\documentclass[]{spie}  
\usepackage[]{graphicx}
\def\spider{{\sc Spider }}

\usepackage{color}

\title{Thermal Architecture for the SPIDER flight cryostat} 

\author{J.E.~Gudmundsson\supit{a}, P.A.R.~Ade\supit{b}, M.~Amiri\supit{c}, S.J.~Benton\supit{d}, R.~Bihary\supit{e}, J.J.~Bock\supit{f,g}, J.R.~Bond\supit{h}, J.A.~Bonetti\supit{f}, S.A.~Bryan\supit{e}, H.C.~Chiang\supit{a}, C.R.~Contaldi\supit{i}, B.P.~Crill\supit{f,g}, D.~O'Dea\supit{i}, M.~Farhang\supit{d}, J.~Filippini\supit{g}, L.M.~Fissel\supit{d}, N.N.~Gandilo\supit{d}, S.~Golwala\supit{g}, M.~Hasselfield\supit{c}, M.~Halpern\supit{c}, K.R.~Helson\supit{e}, G.~Hilton\supit{j}, W.~Holmes\supit{f}, V.V.~Hristov\supit{g}, K.D.~Irwin\supit{j}, W.C.~Jones\supit{a}, C.L~Kuo\supit{k}, C.J.~MacTavish\supit{l}, P.~Mason\supit{g}, T.E.~Montroy\supit{e}, T.~Morford\supit{g}, C.B.~Netterfield\supit{d}, A.S.~Rahlin\supit{a}, C.D.~Reintsema\supit{j}, J.E.~Ruhl\supit{e}, M.C.~Runyan\supit{g}, M.~Schenker\supit{g}, J.A.~Shariff\supit{d}, J.D.~Soler\supit{d}, A.~Trangsrud\supit{g}, R.~Tucker\supit{g}, C.~Tucker\supit{b}, and A.~Turner\supit{f}
\skiplinehalf
\supit{a}Department of Physics, Princeton University, Princeton, NJ, USA; \\
\supit{b}School of Physics and Astronomy, Cardiff University, Cardiff,
UK; \\
\supit{c}Department of Physics and Astronomy, University of British
Columbia, Vancouver, BC, Canada; \\
\supit{d}Department of Physics, University of Toronto, Toronto, ON,
Canada; \\
\supit{e}Department of Physics, Case Western Reserve University,
Cleveland, OH, USA; \\
\supit{f}Jet Propulsion Laboratory, Pasadena, CA, USA; \\
\supit{g}Department of Physics, California Institute of Technology,
Pasadena, CA, USA; \\
\supit{h}Canadian Institute for Theoretical Astrophysics, University
of Toronto, Toronto, ON, Canada; \\
\supit{i}Department of Physics, Imperial College, University of
London, London, England; \\
\supit{j}National Institute of Standards and Technology, Boulder, CO, USA; \\
\supit{k}Department of Physics, Stanford University, Stanford, CA, USA; \\
\supit{l}Kavli Institute for Cosmology, University of Cambridge, Cambridge, UK }

\authorinfo{Send correspondence to Jon E. Gudmundsson: E-mail: jgudmund@princeton.edu, Tel: 1 609 216 4484}

\begin{document} 
\maketitle 

\begin{abstract}
We describe the cryogenic system for {\sc spider}, a balloon-borne microwave polarimeter that will map $8\%$ of the sky with degree-scale angular resolution. The system consists of a $1284 \:\mathrm{L}$ liquid helium cryostat and a $16\:\mathrm{L}$ capillary-filled superfluid helium tank, which provide base operating temperatures of 4 K and 1.5 K, respectively. Closed-cycle $^3\mathrm{He}$ adsorption refrigerators supply sub-Kelvin cooling power to multiple focal planes, which are housed in monochromatic telescope inserts. The main helium tank is suspended inside the vacuum vessel with thermally insulating fiberglass flexures, and shielded from thermal radiation by a combination of two vapor cooled shields and multi-layer insulation. This system allows for an extremely low instrumental background and a hold time in excess of 25 days. The total mass of the cryogenic system, including cryogens, is approximately $1000 \:\mathrm{kg}$. This enables conventional long duration balloon flights. We will discuss the design, thermal analysis, and qualification of the cryogenic system. 
\end{abstract}


\section{INTRODUCTION}
\label{sec:intro}  

Inflation successfully resolves the horizon and flatness problems, and predicts the superhorizon modes that are observed in the Cosmic Microwave Background (CMB) anisotropies. The simplest GUT scale theories of inflation naturally produce primordial gravitational waves that lead to tensor perturbations affecting the development of quadrupolar anisotropies at the surface of last scattering. These anisotropies in turn imprint a small gradient-free polarization signal in the CMB.\cite{Guth1981,Steinhardt1982,Baumann2009} This relic signal, referred to as the B-mode, has yet to be measured, although great progress has been made toward improving instrument sensitivities.\cite{Hunt2003} A number of CMB experiments are being developed to search for the B-mode signal, as its detection would provide extremely strong evidence for inflation.\cite{Oxley2004,Hileman2009,Nguyen2008} These experiments are designed to push the current B-mode sensitivity down by an order of magnitude.\cite{Chiang2010}

\spider is a balloon-borne large-scale CMB polarimeter.\cite{Filippini2010} Through direct observation of the $I$, $Q$, and $U$ Stokes parameters, it will map CMB polarization over about $8\%$ of the sky at degree scale resolution, producing power spectra with multipole moments up to $\ell \approx 300$. The \spider flight cryostat is designed to maintain multiple monochromatic mm-wavelength telescopes\cite{Runyan2010} at sub-Kelvin temperatures for longer than 25 days. This facilitates long duration ballooning flights, which are known to last up to 42 days.\cite{Seo2008} 

\section{Design}

The \spider science goals require cooling of six 1.3~m long telescope inserts with 30~cm apertures down to 1.5~K. The optical throughput and physical dimension of the science instruments determine the scale of the cryostat as well as the minimum parasitic load to the helium bath of several hundred milliwatts.  The enthalpy of the helium vapor produced by this load provides enough cooling power at roughly 20 and 110~K to eliminate the need for a separate liquid nitrogen bath. This helium-only system also reduces the total mass and simplifies the design without compromising thermal performance. See Tab.\ \ref{tab:specs} for an account of main characteristics.

\begin{figure}[t!]
   \begin{center}
   \begin{tabular}{c}
   \includegraphics[width = 0.98\textwidth]{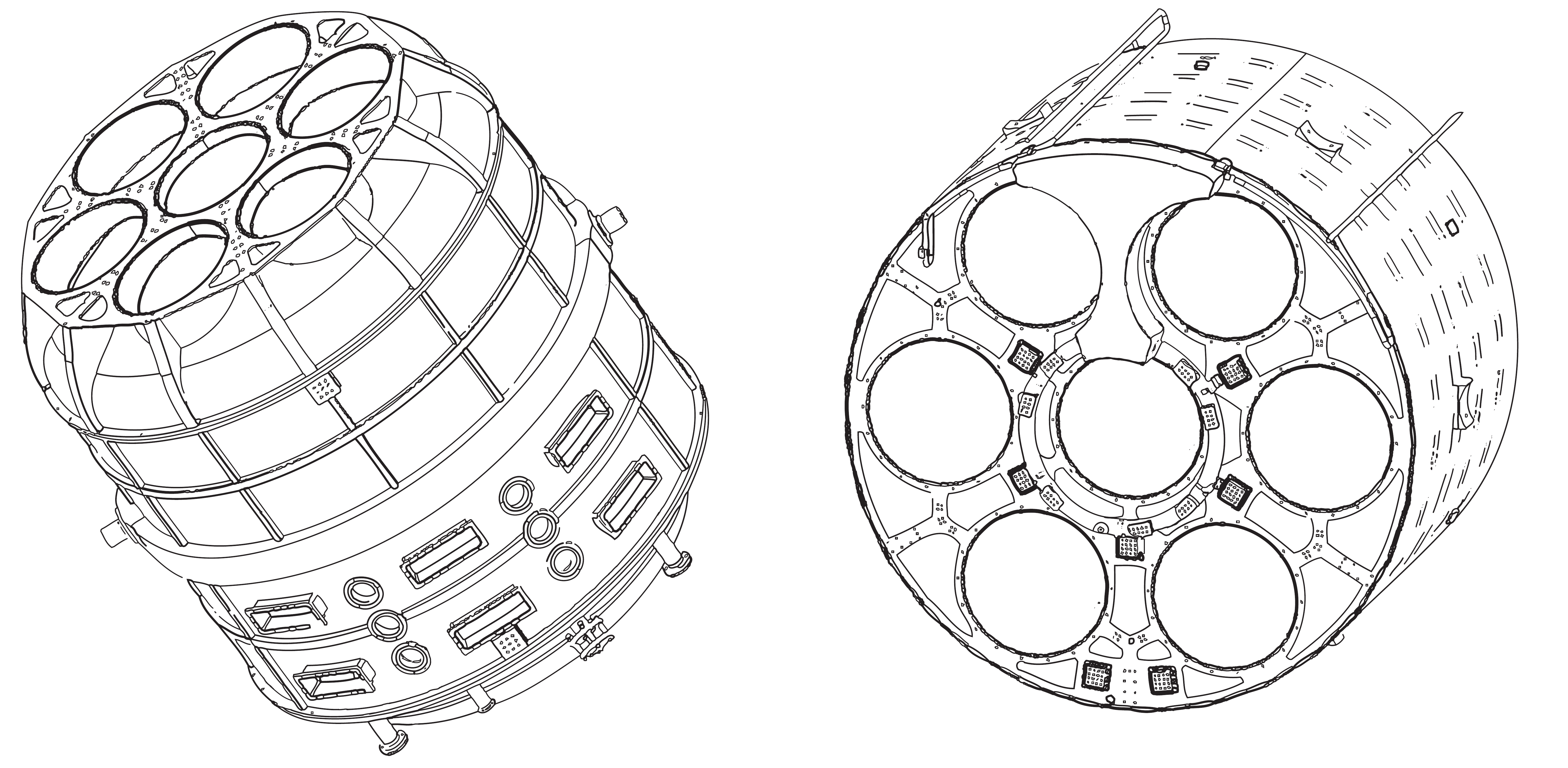}   
   \end{tabular}
   \end{center}
   \caption[example] 
   { \label{fig:VVandMTSFT} 
\emph{Left:} CAD model of the vacuum vessel, the outermost stage of the flight cryostat. The dry weight of the cryogenic assembly excluding telescope inserts is $850 \:\mathrm{kg}$. All fill and vent lines exit the VV at the bottom. Ports on the cylinder sides provide hermetic connections to housekeeping electronics. Two trunnions on the center portion of the VV walls attach to the elevation drive on the gondola. Current observation plans do not anticipate the use of the center insert. A cross-section through the VV can be seen in Fig.\ \ref{fig:schematic}. \emph{Right:} CAD model of the main tank and superfluid tank assemblies. Note the seven telescope inserts constituting extruded cuts through the MT. The SFT has a ring-like structure that connects to another larger cryogenic volume located below the MT. Explosion-bonded thermal contact areas both on the MT and SFT are strapped to each insert to provide cooling power directly from the $4 \:\mathrm{K}$ and $1.5 \:\mathrm{K}$ baths. The copper straps used are not shown on this schematic. The cylindrical MT is $1.69 \:\mathrm{m}$ in diameter, $1.14 \:\mathrm{m}$ long, and weighs $220 \:\mathrm{kg}$. The insert diameter is $419 \:\mathrm{mm}$ and the thickness of the MT walls varies between $4$ and $6 \:\mathrm{mm}$. Observations will be performed with the cryostat tilted at $40^\circ$ elevation such that the bulk of the SFT is above the ring like structure, which holds only about $0.5 \:\mathrm{L}$.}
   \end{figure}

The cylindrically shaped flight cryostat has five main components that are illustrated in Figs.\ \ref{fig:VVandMTSFT} and \ref{fig:schematic}. Starting from the inside, the components are named: Superfluid Tank (SFT), Main Tank (MT), Vapor Cooled Shields 1 and 2 (VCS1, VCS2), and Vacuum Vessel (VV). The bulk of the cryostat is made out of aluminum 1100 (VCS1, VCS2), chosen for its high thermal conductivity, and aluminum 5083 (MT, SFT), which maintains its strength after welding. The cryogenic assembly consists of a cylindrical $1284 \:\mathrm{L}$ Liquid Helium (LHe) main tank, connected through a capillary tube to a $16 \:\mathrm{L}$ superfluid tank. Prior to launch, the pressure in the SFT will be reduced to hundreds of Pascals, making the liquid superfluid, and then capped off. This will reduce the temperature of the SFT down to approximately $1.5 \:\mathrm{K}$, which is a suitable base temperature for the 10 STP-liter closed-cycle $^3\mathrm{He}$ adsorption refrigerators that cool each focal plane. VCS1 surrounds both tanks and serves as a radiation shield from warmer stages, while intercepting conduction and accommodating filters, which need to be maintained at low temperatures to reduce in-band parasitics. VCS2 provides additional radiation shielding from the VV, which is coupled to ambient temperatures. The dry weight of the cryogenic assembly is roughly $850 \:\mathrm{kg}$. 

\begin{table}[h]
\caption{Flight cryostat main specifications.} 
\label{tab:specs}
\begin{center}       
\begin{tabular}{|l|l|} 
\hline
\rule[-1ex]{0pt}{3.5ex}  Vacuum Vessel height & $2.43 \:\mathrm{m}$ \\
\hline
\rule[-1ex]{0pt}{3.5ex}  Vacuum Vessel diameter & $2.05 \:\mathrm{m}$  \\
\hline
\rule[-1ex]{0pt}{3.5ex}  Vacuum Vessel volume & $5.7 \:\mathrm{m^3}$  \\
\hline
\rule[-1ex]{0pt}{3.5ex}  Main Tank net cryogenic volume & $1284 \:\mathrm{L}$ \\
\hline
\rule[-1ex]{0pt}{3.5ex}  Superfluid Tank net cryogenic volume & $16 \:\mathrm{L}$ \\
\hline
\rule[-1ex]{0pt}{3.5ex}  Mass of cryogenic assembly & $850 \:\mathrm{kg}$ \\
\hline
\rule[-1ex]{0pt}{3.5ex}  Hold time & $\geq 25$ days  \\
\hline
\end{tabular}
\end{center}
\end{table} 

\subsection{Performance Requirements} 
\label{sec:PerfReq}
The \spider cryogenic system is considerably larger than comparable systems on other balloon borne CMB experiments \cite{Masi1999,Oxley2004,Silverberg2005}. The cryostat was designed to fulfill the following performance requirements: Have a minimum hold time of 25 days, while sustaining non-parasitic instrument heat loads of  $12 \:\mathrm{mW}$, $550 \:\mathrm{mW}$, $4 \:\mathrm{W}$, and $9 \:\mathrm{W}$ to the SFT, the MT, VCS1, and VCS2, respectively; and to have VCS1 operating at roughly $20 \:\mathrm{K}$ and VCS2 at temperatures lower than $110 \:\mathrm{K}$. The heat loads to VCS2, VCS1, and the MT are mostly due to optical loading. The \spider Half-Wave Plates\cite{Bryan2010} (HWP), located on top of the MT, are also likely to cause considerable loading during operation. A total cooling power of $12 \:\mathrm{mW}$ needs to be supplied by the SFT to six closed-cycle $^3\mathrm{He}$ adsorption refrigerators. Similar thermal behavior has been observed in a smaller test cryostat built to mimic the design of the \spider flight cryostat. This smaller cryostat enables testing of a single telescope insert and reduces the amount of resources needed for various forms of characterization.

\subsection{Structural Requirements}

A strict mass budget needs to be maintained due to constraints set by ballooning.\cite{BillStepp} \spider will have a total flight mass of 3200--3600 kg. This includes a roughly $1400 \:\mathrm{kg}$ flight cryostat with cryogens and science instruments, $110 \:\mathrm{kg}$ gondola, $600 \:\mathrm{kg}$ power systems, $450 \:\mathrm{kg}$ ballast, and a $450 \:\mathrm{kg}$ flight train. The total scientific mass is designed not to exceed $2300 \:\mathrm{kg}$.

Finite Element Analysis (FEA) shows that the design meets the following structural requirements:
\begin{itemize}
	\item Lowest resonance frequency of flight cryostat above $15 \:\mathrm{Hz}$ 
	\item Flight cryostat able to withstand $10 \:g$ parachute shock	
	\item Torsional spring constant of MT with respect to VV greater than $2 \:\mathrm{Nm/ \mu rad}$ 
	\item Radial spring constant of MT with respect to VV greater than $10 \:\mathrm{kN/mm}$	
\end{itemize}

\begin{figure}[t!]
\begin{center}
\includegraphics[width = 0.98\textwidth]{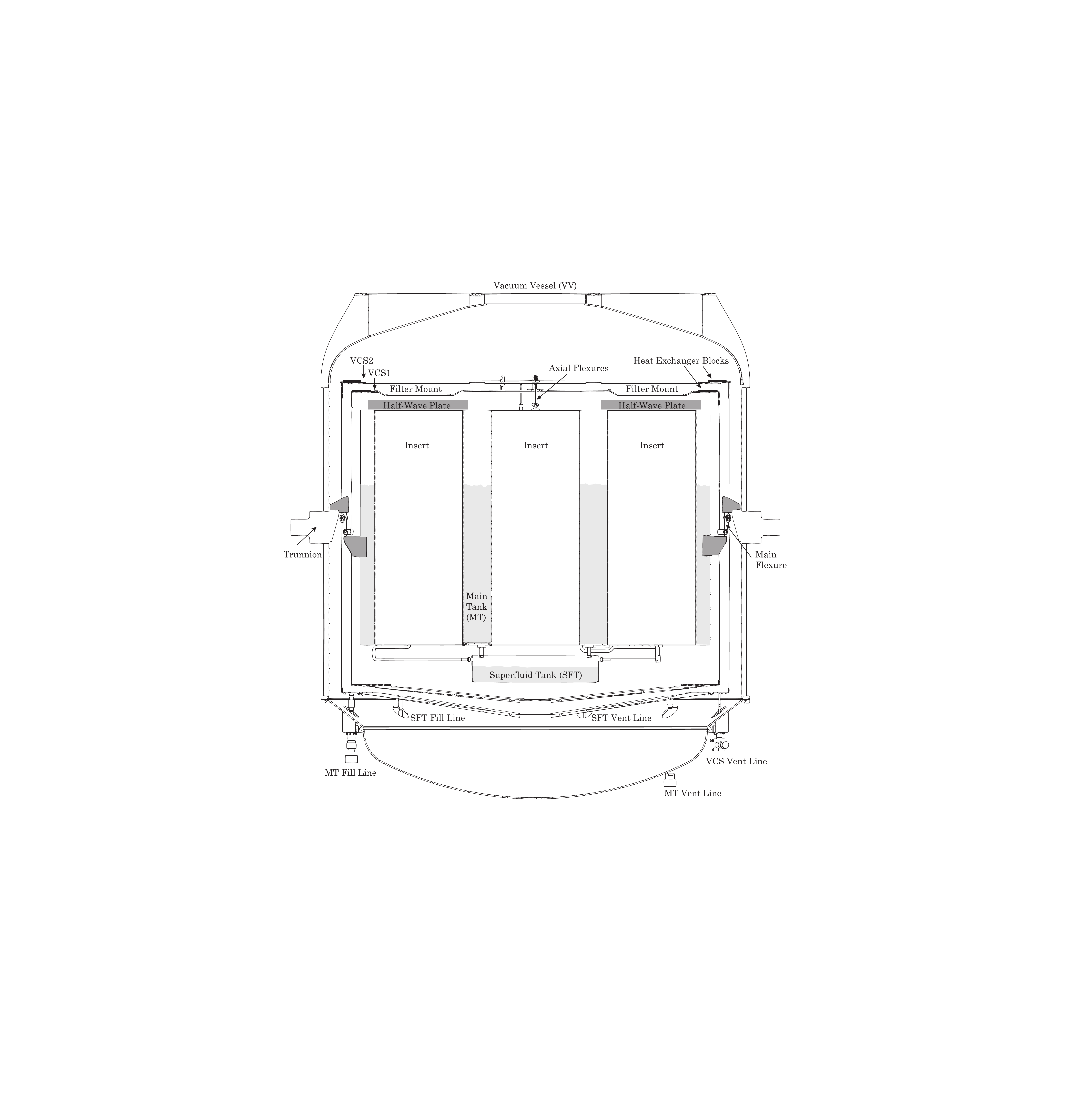}
\end{center}
\caption{\textsl{Cross section through the flight cryostat based on a CAD model and drawn to scale. A number of components have been suppressed to simplify the schematic. Shown are the five different stages of the flight cryostat, namely, the vacuum vessel, VCS2, VCS1, the main tank, and the superfluid tank. Parts of the SFT fill and vent lines are shown as they connect to the SFT. All cryogenic ports are shown as they exit the cryostat at the bottom of the VV. The main flexures are heat sunk at VCS1 and VCS2. Heat exchanger blocks are located on the upper perimeter of both VCS1 and VCS2. Three unpopulated inserts can be seen in this schematic. The locations of the half-wave plates are marked. The components that have been suppressed include kevlar tensioned snubber blocks and MLI located on the outsides of VCS1 and VCS2. During observation the flight cryostat will be nominally tilted at $40^\circ$ elevation. The trunnions shown in this schematic attach to an elevation drive on the gondola.}}
\label{fig:schematic}
\end{figure}

The requirement on minimum resonance frequencies is set by the detector signal bandwidth. The ability to withstand the $10 \:\mathrm{g}$ parachute shock sustained during flight termination is important for successful recovery and turnaround of scientific equipment. Requirements on torsional and radial spring constants are set by pointing constraints and concerns regarding static deflections. Pointing measurements and scan modulations are facilitated by the \spider gondola.

Kevlar-tensioned snubber blocks make up a suspension system designed to minimize translation of the main tank with respect to the vacuum vessel. More than a dozen snubber blocks are anchored on the walls of the MT and connected with Kevlar ties to the inside of the VV. This suspension system is especially important for shock absorption during landing. 

The \spider gondola is designed to preserve the structural integrity of the experiment under flight conditions, using a minimum fraction of the payload mass limit. It is constructed with carbon fiber reinforced polymer tubes attached to aluminum inserts using an adhesive. The aluminum inserts are attached together with bolts in aluminum multitube nodes. The balloon-flight conditions include $10 \:g$ vertical acceleration during the parachute shock and large angular acceleration on the inner frame during launch. Each component of the gondola was designed and simulated under these scenarios using FEA to optimize for strength and mass. The results of the analysis were validated by mechanical testing of both the adhesive and the bolt junctions. The final structure comprises only 3\% of the total flight mass.

\subsection{Radiation Shields, Heat Exchangers, and Multi-layer Insulation}

Two intermediate vapor cooled radiation shields, VCS1 and VCS2, made out of aluminum 1100, serve as thermal anchors for multi-layer insulation, filter blocks, and heat exchangers. VCS1 is supported by the MT, while VCS2 is supported from the VV. Aluminum 1100 is a relatively conductive alloy, therefore the cooling power from the heat exchangers is distributed more effectively, leading to improved isothermality of the vapor cooled shields.

Six compact heat exchangers are symmetrically placed on the top sides of both VCS1 and VCS2, see Figs.\ \ref{fig:schematic} and \ref{fig:FlexAndHx}. Cryogenic boil-off is forced to go through these flow-restrictive heat exchangers, cooling the respective stages, and thus providing negative feedback.\cite{Kays, deWitt} The heat exchangers are made of stainless steel blocks enclosing densely packed horizontal mesh (VCS1) or pellets (VCS2), both made of copper. Stainless steel was chosen for ease of welding. The heat exchangers also provide cooling power to the infrared-blocking filters and are thermally tied to the filter mount points with copper straps. The filter mounts should be as cold as possible to reduce optical background on the detectors. 

The outer sides of VCS1 and VCS2 are layered with aluminized polyester films (known as aluminized Mylar, MLI for Multi-layer Insulation, or superinsulation): 16 layers on the outside of VCS1 and 52 layers on the outside of VCS2. These $6.4 \:\mathrm{\mu m}$ thin layers of thermoplastic polymer provide a lightweight substrate for highly reflective $35 \:\mathrm{nm}$ thick aluminum layers on both sides. The MLI is designed to provide sufficient radiation damping to maintain large temperature gradients between stages. A $0.1 \:\mathrm{mm}$ thick, spun-bound polyester is placed between each layer of Mylar to reduce thermal conduction, which has the potential to dominate radiative effects if the layers have strong thermal contacts. The surface area of each stage ranges between 10--14 $\mathrm{m}^2$. The MLI packaging density of $14 \:\mathrm{layers/cm}$ reduces compression that would otherwise lead to undesirable conductance effects. This is roughly a factor of two lower than the quoted optimal packaging density for MLI.\cite{Bapat1990,Eyssa1978} However, lower packaging densities enable successful evacuation of the vacuum vessel.

\subsection{Flexures and Plumbing}

\begin{figure}[t]
   \begin{center}
   \begin{tabular}{c}
   \includegraphics[width = 0.80\textwidth]{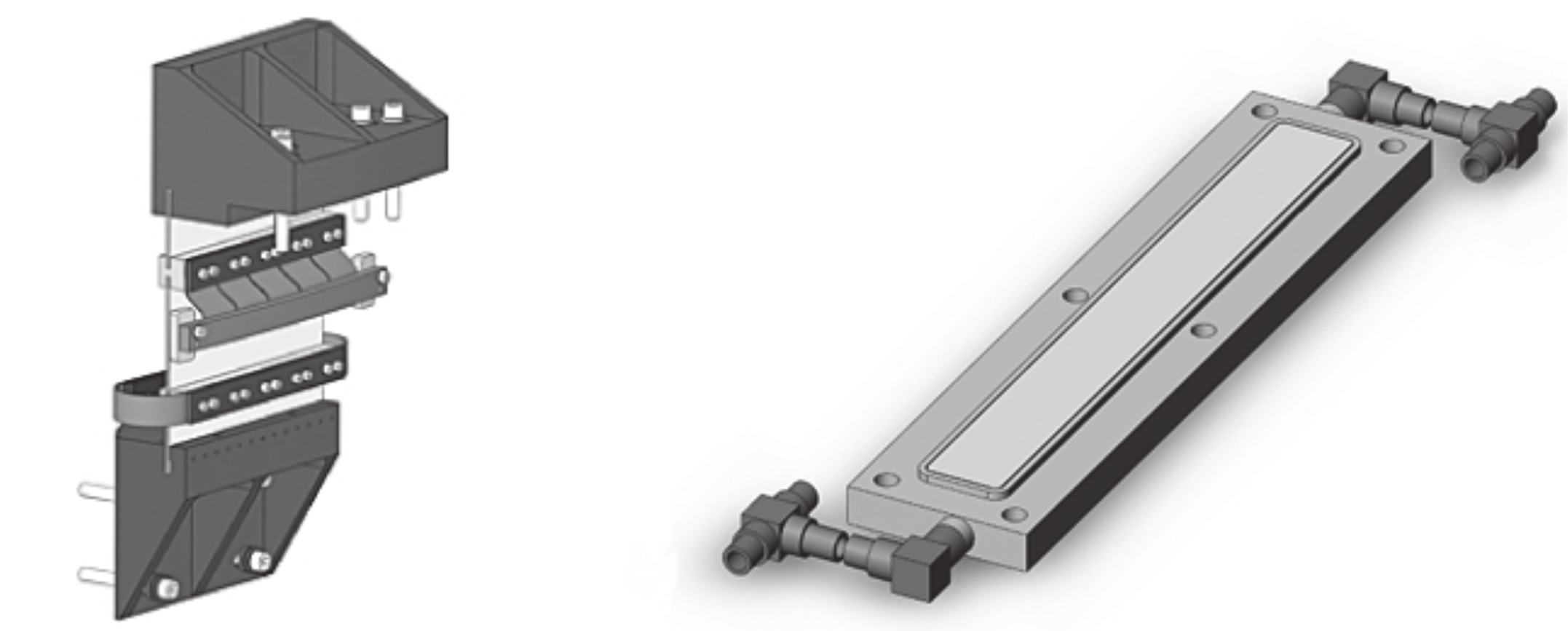}
   \end{tabular}
   \end{center}
   \caption[] 
   { \label{fig:FlexAndHx} 
\emph{Left:} One of the six main structural flexures which are symmetrically spaced around the main tank cylinder. These flexures support the MT off of the VV through a G-10 sheet which is thermally damped by copper heat straps connected to VCS1 and VCS2. The G-10 flexures are $1.6 \:\mathrm{mm}$ thick, $127 \:\mathrm{mm}$ wide, and $114 \:\mathrm{mm}$ long. \emph{Right:} One of the six heat exchangers placed on the top of VCS2. The $15 \:\mathrm{cm}$ long rectangular compartment is randomly filled with copper pellets creating high flow impedance, which extracts enthalpy from the helium boil-off. The heat exchanger system forms a double annular structure on each of the vapor cooled shields. Helium vapor exiting the cryostat through the VCS vent must go through one of the six heat exchangers on both VCS1 and VCS2. The cylindrical symmetry of the heat exchanger system ensures that equal enthalpy is extracted by the heat exchangers.}
\end{figure}

The main tank is supported by the vacuum vessel through six G-10/Aluminum flexures symmetrically placed on the cylinder sides, see Fig.\ \ref{fig:FlexAndHx}. G-10, a lightweight, high tensile strength fiberglass, is an extremely poor conductor of heat, and therefore ideal for cryogenic flexures.\cite{Runyan2008} In order to dampen heat flow from the VV to the MT the flexures are heat sunk at VCS1 and VCS2 using copper straps. Similar but considerably smaller G-10 flexures, $33 \:\mathrm{mm}$ long, $29 \:\mathrm{mm}$ wide, and $0.8 \:\mathrm{mm}$ thick, connect VCS1 to the MT and VCS2 to the VV at six points on the cylinder sides. Three axial flexures help align VCS1 with respect to the MT, and six flexures connect the SFT to the bottom of the MT. 

\spider will scan at a $40^\circ$ elevation such that the liquid level of a normal liquid will not lie in the plane defined by the bottom of the main tank. For this reason, and since the superfluid tank will never be entirely full, it is important that the helium in the SFT is superfluid to ensure that sufficient cooling power is supplied to each insert. This fact motivates the shape of the SFT, which can be seen in Fig. \ref{fig:VVandMTSFT}, and the placement of thermal contact areas on the ring-like structure. Explosion-bonded Al/Cu transition plates provide reliable thermal contact areas from both the MT and the SFT. Parts of the aluminum are milled out such that the copper is in direct contact with the cryogen. The remaining aluminum is then welded on to the corresponding tank. The transition plates are connected to custom-made copper heat straps, and supply the necessary cooling power to telescope inserts and sub-Kelving cooling stages. The MT thermal contact areas can also be seen in Fig \ref{fig:VVandMTSFT}.

Plumbing lines are made out of type 304 stainless steel, due to its low thermal conductivity and suitability for welding. There are five plumbing lines leading from the outside of the VV to either the MT or the SFT. These are: The MT fill and vent lines which have a $3/4$ inch Outer Diameter (OD), the SFT fill and vent lines which have a $1/2$ inch OD, and the VCS vent line which has a $1/4$ inch OD. Aluminum/stainless transitions are made from explosion-bonded blocks that are welded in place. MT fill and vent lines will be capped off prior to launch, forcing helium boil-off to leave the MT through the heat exchangers. All plumbing lines, excluding the VCS vent line, are strategically heat sunk at VCS2 and not at VCS1. The length of the MT/SFT vent and fill lines is approximately $2.7 \:\mathrm{m}$ in all cases, while the average travel of gas through the VCS vent line system is about 12~m. The size of the VCS vent line system leads to high heat exchanger efficiency, see Sec.\ \ref{sec:conduction}. Vent lines are positioned on cryogenic tanks such that boil-off will be able to exit the cryostat when it is tilted at a $40^\circ$ elevation angle and full of cryogen. The expected helium flow rate during operation is roughly $20 \:\mathrm{SLPM}$.
  
   \begin{figure}
   \begin{center}
   \begin{tabular}{c}
   \includegraphics[width = 0.98\textwidth]{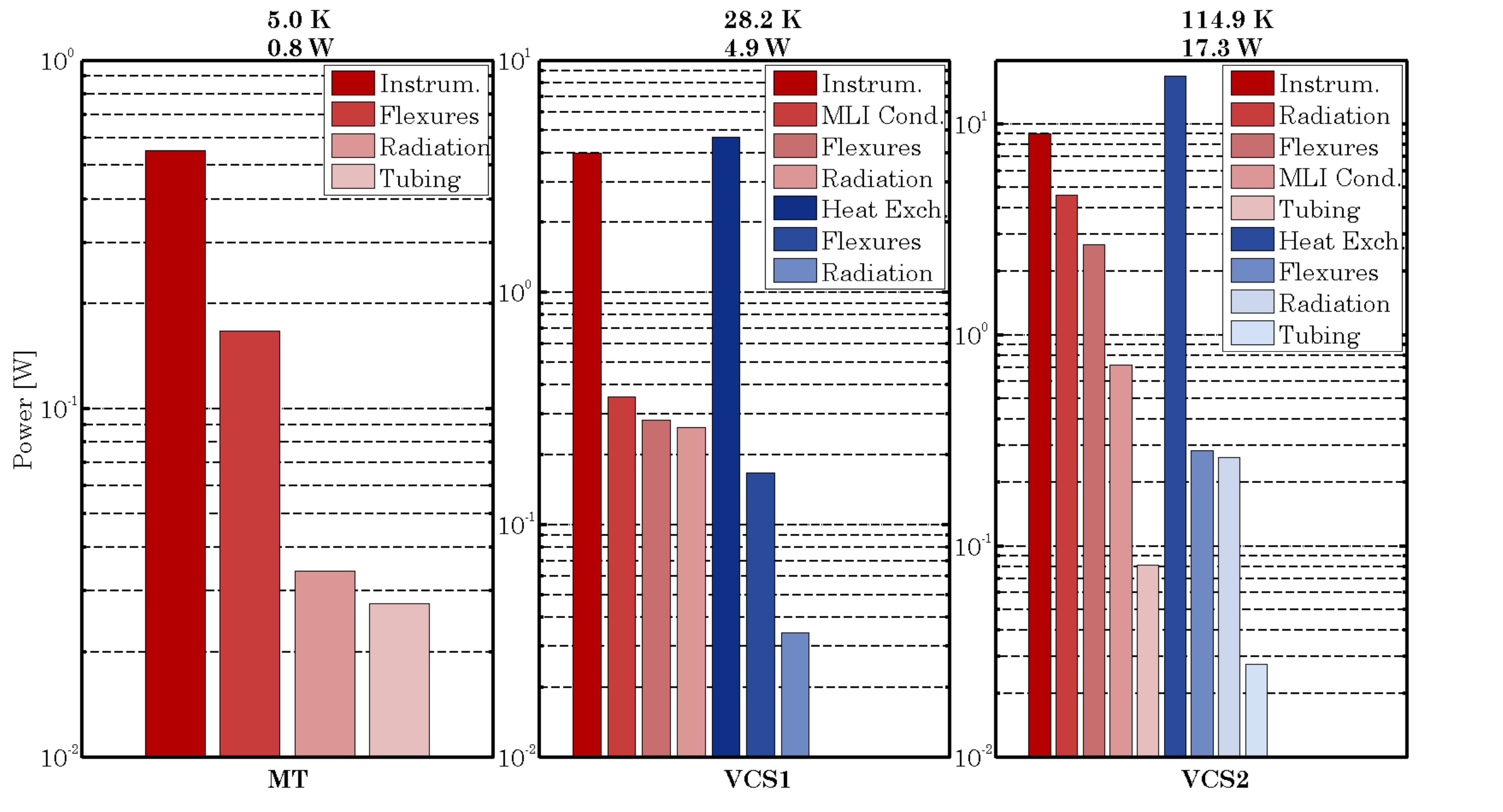}
   \end{tabular}
   \end{center}
   \caption[] 
   { \label{fig:budget} The predicted thermal budget for the main tank, VCS1, and VCS2, with expected instrument loading. The net heat load to VCS1 and VCS2 is zero at equilibrium. Red bars represent positive heat input while blue bars correspond to negative heat input (cooling). The text above each graph shows the average temperature of the respective stage, along with the total positive heat input. Note that instrument loading dominates the heat load to all stages. Apart from instrument loading, conduction through flexures, MLI conduction, and radiation are thought to contribute the largest heat load to the MT, VCS1, and VCS2 respectively. The main cooling power to VCS1 and VCS2 is provided by the heat exchangers. For the graphs associated with VCS1 and VCS2 cooling terms begin with cooling from heat exchangers and continue to the right.}
   \end{figure} 

\section{Thermal Analysis}

A numerical model was constructed to account for various forms of heat transfer in the cryogenic system. The goal was to create the simplest model that would still capture observations. Data from cryogenic test runs were then used to optimize the predictive power of the model, leading to improved understanding of thermal behavior. A chi-squared significance test was performed, which suggested that the model captured the thermal behavior over a range of observations without over-parametrization.

Heat transfer in this cryogenic system is attributed to either radiation or conduction, since these two processes should account for nearly all heat transfer. The numerical model divides the system into five stages assumed to be isothermal. These are the VV, VCS2, VCS1, the MT, and the SFT. Care is taken to account for as many forms of heat transfer between stages as possible. Material thermal properties and free parameters are input to the model, which searches for equilibrium temperatures such that the net heat load to both vapor cooled stages is zero.

\subsection{Conduction}
\label{sec:conduction}

Thermal conduction between two points in a linear isotropic medium with a given cross-sectional area $A$ and length $L$ can be calculated if the thermal conduction, $k$, is known as a function of temperature:
\begin{equation}
\dot{Q} = \frac{A}{L}\int\limits _{\mathrm{T_1}}^{\mathrm{T_2}} k(T)dT.
\end{equation}
The model evaluates this integral for various materials connecting different stages, including stainless steel, phosphor bronze, aluminum alloys, and G-10. A considerable fraction of the thermal budget is attributed to MLI conduction. Due to complex geometries, uncertainties about compressive loads, etc.\ it becomes difficult to estimate the MLI thermal conductance as a function of temperature. Instead, somewhat empirical estimates are adopted.

The heat input to the main tank goes into boiling off cryogenics such that a steady gas flow out of the MT will be established at equilibrium. This gas is forced to go through the heat exchangers at VCS1 and VCS2, providing negative feedback. The cooling power supplied by the heat exchangers on VCS1 can be written as
\begin{equation}
\dot{Q} = \dot{m}\eta\int\limits _{T_\mathrm{MT}}^{T_\mathrm{VCS1}}C_p(T)dT,
\end{equation}
where $\dot{m}$ is the mass flow through the heat exchangers, $\eta$ is the heat exchanger's efficiency,  $C_p$ is the specific heat of the gas, and $T_\mathrm{MT}$ and $T_\mathrm{VCS1}$ are the mean temperatures of the MT and VCS1 respectively. The cooling power to VCS2 can be written in a similar manner. Comparison of thermal performance with predictions of the thermal model suggest that the heat exchangers are operating with efficiencies close to unity. This is not uncommon.\cite{Steffensrud1991,Okamoto2001}

The flight cryostat has MLI installed in such a way that reduces compression, which would otherwise increase conduction through the insulation, and enables proper evacuation of interstitial gas.\cite{Bapat1990,Eyssa1978,Jacob1992a} However, thermal conduction is inevitable when adjacent layers are in sporadic contact. In the absence of a leak, the pressure inside the VV remains roughly constant after equilibrium has been reached. Gas particles that have not condensed will then conduct heat between stages at a steady rate. If the pressure is sufficiently low the heat conduction through the MLI due to rarefied gas is negligible. Thermal conduction is then due to electron and phonon propagation. As the rarefied gas pressure increases, there comes a point when gas conductance becomes non-negligible. This can be due to outgassing and poor evacuation of MLI layers. MLI conduction is then roughly linearly proportional to gas number densities. At this point, gas is still in the free-molecular regime, although it is starting to contribute significantly to heat transfer.\cite{Corruccini1959,Sun2009} This behavior can also be seen in cryogenic systems that have microscopic leaks. Small leaks can be negated by the installation of activated charcoal or zeolite adsorbers.\cite{Kumita1995,Duband1987} 

The thermal model has free parameters which describes the effective thermal conductance, excluding radiative transfer, through the MLI. Thermal characterization tests have helped improve this estimate. It is assumed that the conductivity through MLI is roughly $k_\mathrm{MLI} = 0.5 \:\mathrm{\mu W/cmK}$, which is comparable to observed values from other experiments.\cite{Scurlock1976} 
\subsection{Radiation}

Estimating radiative coupling between various stages is difficult due to the complex structure of the \spider flight cryostat. No attempts have been made to calculate view factors for the intricate cylindrical geometries inside the flight cryostat. Instead, different stages are assumed to couple radiatively in the same way as parallel planes. This approximation becomes more accurate when the physical separation between stages is small. The estimated effective area of each stage then becomes a free parameter incorporating properties such as view factors. This simplification also facilitates model optimization.

The Hagen--Rubens relation, derived from Maxwell's equations, states that the normal spectral emissivity of a conductor is \cite{Bock1995,Heaney}
\begin{equation}
\epsilon = \sqrt{16\pi\varepsilon _0 \rho\nu},
\label{HagRub}
\end{equation}
where $\rho$ is the resistivity of the conductor, $\varepsilon _0$ is the vacuum permittivity, and $\nu$ is the frequency of radiation emitted. It can also be shown that emissivity is angle-dependent, and in the case of conductors, that the emissivity is highest at large angles normal to the plane. In order to simplify calculations, this model uses the total hemispherical emissivity (from here on referred to simply as emissivity), which is the emissivity averaged over frequency and angle. Temperature variation in emissivity is assumed to vary according to a power law and a linear temperature profile through the MLI is assumed as a first approximation.\cite{Bapat1990}

The radiative heat load per unit area between two infinite black body planes, $\epsilon = 1$, at temperatures $T_{\mathrm{C}}$ and $T_{\mathrm{H}}$ is
\begin{equation}
Q = \sigma (T_{\mathrm{H}}^4-T_{\mathrm{C}}^4),
\end {equation}
where $\sigma$ is the Stefan--Boltzmann constant. The net radiative heat transfer per unit area between two parallel infinite plains at temperature $T_\mathrm{H}$ and $T_\mathrm{C}$ with emissivities $\epsilon_{\mathrm{H}}$ and $\epsilon_{\mathrm{C}}$ can be shown to be
\begin{equation}
\dot{Q} =\sigma\left(\frac{\epsilon_{\mathrm{H}}\epsilon_{\mathrm{C}}}{\epsilon_{\mathrm{H}}+\epsilon_{\mathrm{C}}-\epsilon_{\mathrm{H}}\epsilon_{\mathrm{C}}}\right)\left(T_{\mathrm{H}}^4-T_{\mathrm{C}}^4\right).
\label{eq:rad1}
\end{equation}
This is done by summing up infinite contributions due to reflection and absorption of radiation, assuming no transmission. More complex geometries require the calculation of view factors.\cite{deWitt} For $N+1$ parallel infinite surfaces the net radiative heat transfer per unit area can be shown to be
\begin{equation}
\dot{Q} = \sigma\left(\sum_{i=1}^{N}\frac{1}{\epsilon _{i,i+1}}\right)^{-1}(T_{N+1}^4-T_1^4),
\label{eq:rad2}
\end{equation}
where
\begin{equation}
\epsilon _{i,i+1} \equiv \frac{\epsilon_{\mathrm{i}}\epsilon_{\mathrm{i+1}}}{\epsilon_{\mathrm{i}}+\epsilon_{\mathrm{i+1}}-\epsilon_{\mathrm{i}}\epsilon_{\mathrm{i+1}}}
\label{eq:rad3}
\end{equation}
is the effective emissivity between layers $i$ and $i+1$, having emissivities $\epsilon_{\mathrm{i}}$ and $\epsilon_{\mathrm{i+1}}$, and $T_1$ and $T_{N+1}$ are the temperatures of the first and the last layers respectively. Equation (\ref{eq:rad2}) is generally inversely proportional to $N$. Emissivity is temperature-dependent so the temperature of all layers do affect the radiative load. In estimating radiative coupling between the VV, VCS2, and VCS1, Eq.\ (\ref{eq:rad2}) is employed with $N = 52$, and $N = 16$ respectively. No MLI is present between VCS1 and the main tank and so $N = 1$.

\section{Cryogenic Performance}

A number of cryogenic tests have been performed to characterize the flight cryostat. These tests are designed not only to imitate the expected in-flight loading described in Sec.\ \ref{sec:PerfReq} but also to assess distinct cryogenic characteristics. Many of those tests used liquid nitrogen instead of helium to improve characterization of radiative coupling. These tests are performed using heaters placed at various locations inside the flight cryostat. A grid of thermometers allows monitoring of temperature gradients and enables informed measurements of cryogenic performance. Cryogenic qualification tests are currently being performed to verify that the cryostat is performing as expected at LHe temperatures. Preliminary results suggest that a 25 day hold is achieved.

A closed-cycle $^3\mathrm{He}$ adsorption refrigerator, located inside the flight cryostat, has been cooled to $300 \:\mathrm{mK}$, achieving a hold time of almost 4 days, while only sustaining parasitic loads. Runs performed in the smaller test cryostat show that the closed-cycle $^3\mathrm{He}$ adsorption refrigerators are capable of maintaining a steady base temperature of $300 \:\mathrm{mK}$ with a load of $10 \:\mathrm{\mu W}$ for $3 \:\mathrm{days}$ before a $1 \:\mathrm{hour}$ cycling of the refrigerators is required. The conductance through thermal contact areas connecting the MT and the SFT to the telescope inserts has been measured. These measurements imply that the heat straps are capable of supplying sufficient cooling power to all inserts at a $40^\circ$ elevation. Cryogenic tests simulating transient loads from a HWP show no sign of significant thermal excitations. This suggests that the cryostat is qualified to accommodate up to six half-wave plates without a significant adverse effect. 

\acknowledgments    

The \spider collaboration gratefully acknowledges the support of NASA (grant number NNX07AL64G), the Gordon and Betty Moore Foundation, and NSERC. HCC is supported by a Princeton Fellowship in Experimental Physics. JPF is partially supported by a Moore Postdoctoral Fellowship in Experimental Physics. JEG is partially supported by the ASF Thor Thors Special Contribution Fund. WCJ acknowledges the support of the Alfred P. Sloan Foundation. The \spider cryostat was fabricated by Redstone Aerospace in Longmont
Colorado. We are grateful to Robert Levenduski, Larry Kaylor and Edward
Riedel for their contributions to the project.

The \spider collaboration also extends special gratitude to Andrew E. Lange, who passed away on January 22. Andrew was a leading light of the cosmology community and a driving force behind the Spider project. He is greatly missed by his many collaborators, past and present.

\bibliography{refs}   
\bibliographystyle{spiebib}   

\end{document}